\documentclass[twocolumn,showpacs,preprintnumbers,amsmath,amssymb,prb]{revtex4-1}

\usepackage[english]{babel}
\usepackage{color}
\usepackage[usenames,dvipsnames,svgnames,table]{xcolor}
\usepackage{verbatim}

\usepackage{amsmath}
\usepackage{amssymb}
\newcommand{\bra}[1]{\langle #1|}
\newcommand{\ket}[1]{|#1\rangle}
\newlength{\eqboxstorage}

\usepackage{graphicx}
\usepackage{dcolumn}
\usepackage{bm}

\begin{document}

\title{{\bf k}$\cdot${\bf p} subband structure of the LaAlO$_3$/SrTiO$_3$ interface}

\author{L.W. van Heeringen, G. A. de Wijs, A. McCollam, J.C. Maan and A. Fasolino}

\affiliation {
Radboud University Nijmegen, Institute for Molecules and Materials and High Field Magnet Laboratory , Heyendaalseweg 135, 6525 AJ Nijmegen, The Netherlands
         }


\begin{abstract}
Heterostructures made of transition metal oxides are new tailor-made materials which are attracting  much attention. 
We have constructed a 6-band {\bf k}$\cdot${\bf p} Hamiltonian and used it within the envelope function method to calculate the subband structure of a variety of 
LaAlO$_3$/SrTiO$_3$ heterostructures. 
By use of density functional calculations, 
we determine the {\bf k}$\cdot${\bf p}
parameters describing the conduction band edge of SrTiO$_3$: the three effective mass parameters, $L=0.6104~\text{eV\AA}^2$, $M=9.73~\text{eV\AA}^2$, $N=-1.616~\text{eV\AA}^2$, the spin orbit splitting $\Delta_{SO}=28.5~$meV and the low temperature tetragonal distortion energy splitting $\Delta_T=2.1~$meV.
For confined systems we find  strongly anisotropic non-parabolic subbands. 
As an application we calculate bands, density of states and magnetic energy levels and compare the results  to Shubnikov-de Haas quantum oscillations observed in high magnetic fields.
For typical heterostructures we find that electric field strength at the interface of 
$F = 0.1$~meV/\AA{} for a carrier density of $7.2 \times 10^{12}$~cm$^{-2}$ results in a subband structure that is  similar to experimental results.
\end{abstract}
\maketitle

\section{Introduction}
Oxides of the transition metals, like LaAlO$_3$ and SrTiO$_3$, are intriguing and useful materials, displaying many properties attributed to electron correlations effects like ferro- and antiferromagnetism, colossal magnetoresistance, metal-insulator transitions, and high and low temperature superconductivity.\cite{bookgoodenough} Recently, pulsed laser deposition growth methods succeeded to combine different oxides in heterostructures with layers as thin as 10-20 lattice parameters and relatively sharp interfaces.\cite{ohtomo2002artificial,ohtomo2004high,brinkman2007magnetic} These samples resemble the well known semiconductor heterostructures, since the different band gaps of the two materials and their band line-up at the interface\cite{popovic2008origin,son2009density} can lead to quantization of the electronic states into two-dimensional levels, opening the way to the typical two-dimensional electron systems phenomena.\cite{bastard1988wave}  

The bandstructure of semiconductor heterostructures is quite succesfully described with the effective mass k.p method with wavefunctions that are matched at the interface between adjacent layers. Such a framework is absent for the transition metal oxide heterostructures and is developed in this paper. 
Much interest in these structures has been triggered by the observation in magnetotransport of a high-mobility electron gas at the LaAlO$_3$/SrTiO$_3$ interface,\cite{ohtomo2004high,mannhart2010oxide} an unexpected feature in view of the fact that the constituent materials are both insulators. 
Several models have been proposed to explain the presence of the charge carriers at the interface. A so-called polar catastrophe, namely the charge transfer of half an electron per unit cell ($3.3 \times 10^{14}$ cm$^{-2}$)  at the interface to compensate the  ever increasing electrostatic potential due to the polar layers in LaAlO$_3$, has been often invoked. A wide range of charge densities, also orders of magnitude away from those  expected from this model, have been measured, revealing that details of the structure and in particular the presence of oxygen vacancies also have a crucial role.\cite{popovic2008origin,mannhart2010oxide}

The origin of the charge carriers is not completely understood and there are also conflicting ideas on the nature of the bands that are responsible for the conductivity. 
Recently, tight-binding calculations~\cite{khalsa2012theory} showed that the  bands of the quasi-two-dimensional electron gas are either  atomic-like or delocalized  depending on the carrier density $n$.\cite{note} In the low density regime ($n<1\times10^{14}$ cm$^{-2}$), the electrons are deeply spread into the SrTiO$_3$ due to strong dielectric screening and the subbands are only meV apart. 
For higher densities, nonlinear screening becomes important and the electrons are confined closer to the interface. 

Many experiments and theoretical calculations have been performed for these higher densities showing atomic-like levels, eV apart.\cite{popovic2008origin,santander2011two,delugas2011spontaneous,breitschaft2010twodimensional}. Here, electron correlations clearly play an important role giving rise to enhanced effective masses, effects of localization and Kondo phenomena, that cannot be described in a single particle model.  However many results are reported on high mobility, low density heterostructure samples (densities $n\sim 1\times 10^{13}$ cm$^{-2}$) which show clear Shubnikov de Haas oscillations, with at least one but often several two-dimensional subbands~\cite{mccollam2012quantum,caviglia2010}. Two dimensional magnetotransport with signatures of multiple subband conduction have also been observed in $\delta$ doped SrTiO$_3$.\cite{kozuka1,kozuka2,jalan2010} In the multiple subband case, subband separations of a few meV can be resolved.\cite{mccollam2012quantum,caviglia2010} This low density regime has recently been shown to be related to La-deficient films.\cite{PhysRevLett.110.196804} 

In this paper we focus on the properties of the electrons in the low density regime. In this regime the single particle bandstructure can be effectively described with a 6-band {\bf k}$\cdot${\bf p} approach for the  bands of the bulk materials, and matching of the envelope functions at the interface for the LaAlO$_3$/SrTiO$_3$ heterostructures.  We determined effective mass parameters by fitting the bulk bandstructure as calculated with density functional theory (DFT). 

The effective mass {\bf k}$\cdot${\bf p} method \cite{bastard1988wave} complements  tight-binding  calculations.\cite{zhong2012theory,khalsa2012theory}  in the low density regime where  the electron states are sufficiently extended and where it is more convenient  in view of the large number of atoms in the unit cell. Furthermore, the {\bf k}$\cdot${\bf p} method can easily be extended  to incorporate the effect of  perpendicular and parallel magnetic fields, electric fields and self-consistent calculation if necessary. Our versatile effective mass approach and the parameters that we obtain  can be used in many SrTiO$_3$-based heterostructures,  giving results that are very useful to analyze experiments on these new materials. In particular, the relatively heavy masses experimentally observed in this material system follow  directly from the single-particle band structure.  Our method could be applied to obtain the single particle energy level structure in many of the samples mentioned in previous work, including multiple subband conduction observed on application of an electrostatic potential to a SrTiO$_3$ surface.\cite{ueno2008} 

As an important example, we apply our method to the LaAlO$_3$/SrTiO$_3$ interface which has shown two-dimensional conductivity. We determine the quantized energy levels, the in-plane dispersion and the density of states. Via quasiclassical quantization, we calculate the energy levels in a magnetic field and compare the results to magnetotransport measurements.  Note that when all relevant parameters of the sample are knowns (density, layer thicknesses and doping), there are no free parameters left and the calculation gives the actual energy level structure.  Deviations between theory and experiments should then be attributed to the neglect of correlations in the calculations. Such an accurate theory is therefore an excellent starting point to study correlation effects. 
Our calculations are relevant for all low density SrTiO$_3$-based two-dimensional systems and are  particularly relevant for the analysis of recent multisubband  Shubnikov-de Haas (SdH) oscillations.\cite{mccollam2012quantum} Here, we find anisotropic, non-parabolic bands with quite different in-plane masses and an energy dependent density of states that is important for the interpretation of the SdH experiments, which usually assumes parabolic and isotropic bands. Moreover, as the relevant energy spacings at the interface are meV, incorporation of spin-orbit (SO) coupling is crucially important.  Similar results with small subband separations and heavy in-plane masses are also reported in Refs.~\onlinecite{caviglia2010} and \onlinecite{kozuka2}.

In section \ref{sec:model-bulk}, we model the bulk conduction band structure around the $\Gamma$ point using the {\bf k}$\cdot${\bf p} approach. In section \ref{sec:DFT calculations}, we find the {\bf k}$\cdot${\bf p} parameters by fitting the bulk bands calculated within DFT. We discuss in detail the importance of SO coupling. In section \ref{sec:QWsub}, we use the envelope function method to calculate the subband energy structure of SrTiO$_3$ quantum wells.  In section \ref{sub:SdHcomp}, we introduce an electric field to account for the polar structure of the interface and compare our results to the SdH experiments. Section \ref{sec:S&C} presents a summary of this work.

\section{Bulk band structure model}
\label{sec:model-bulk}
In this section, we construct the {\bf k}$\cdot${\bf p} model of the band structure of the LaAlO$_3$/SrTiO$_3$ interface around the $\Gamma$ point. We first use symmetry arguments to show that the SrTiO$_3$ conduction bands alone are enough to describe the system, making a {\bf k}$\cdot${\bf p} model of LaAlO$_3$ unnecessary. We then model the conduction bands of SrTiO$_3$ with three effective mass parameters $L$, $M$, $N$, the spin orbit splitting $\Delta_{SO}$ and the low temperature tetragonal distortion energy splitting $\Delta_T$. In section~\ref{sec:DFT calculations}, these five {\bf k}$\cdot${\bf p} parameters are determined by fitting to  the DFT bulk band structure of SrTiO$_3$.
\begin{figure}[h]
\begin{center}
\includegraphics[angle=270,width=8cm]{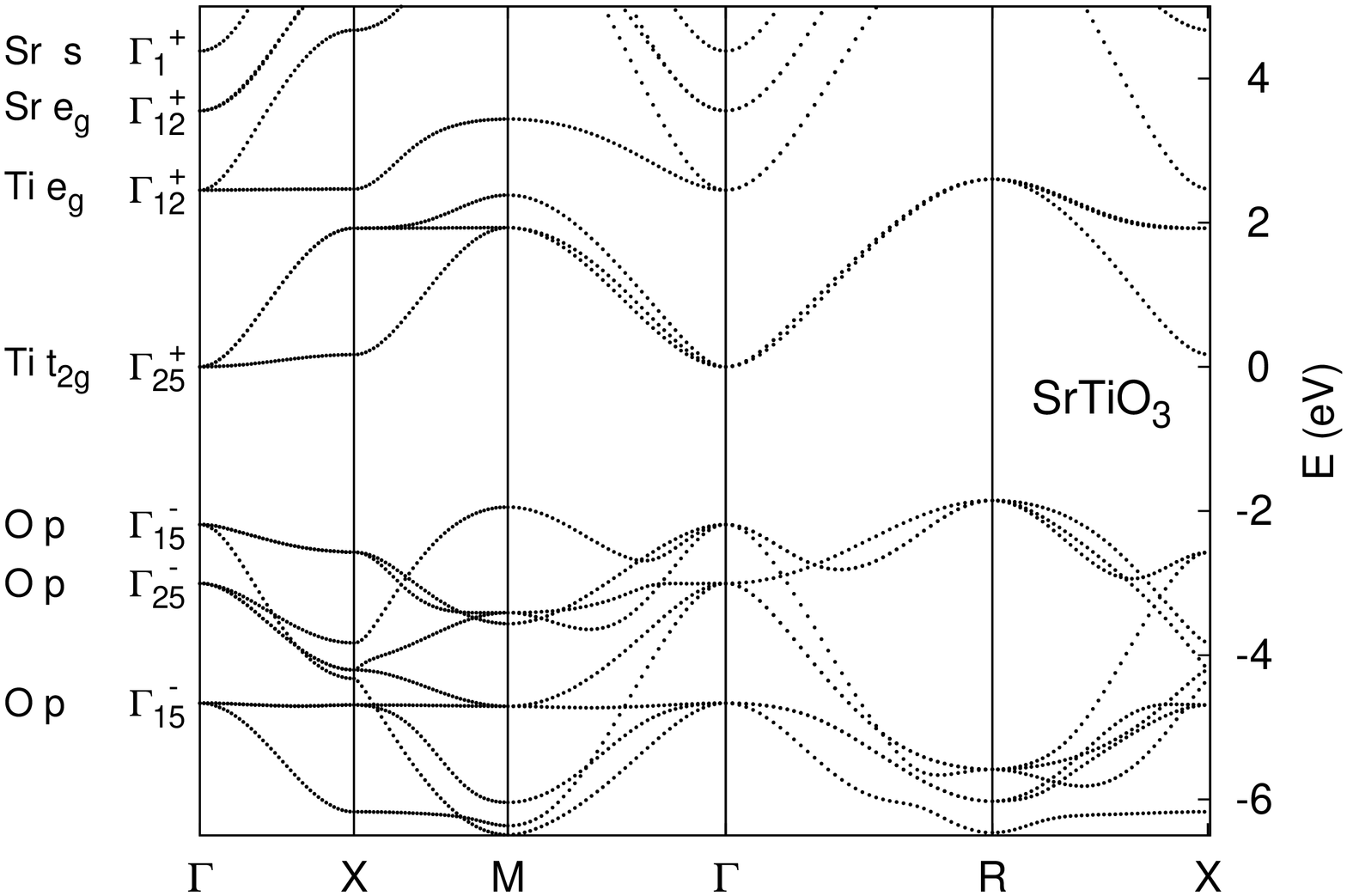}
\includegraphics[angle=270,width=8cm]{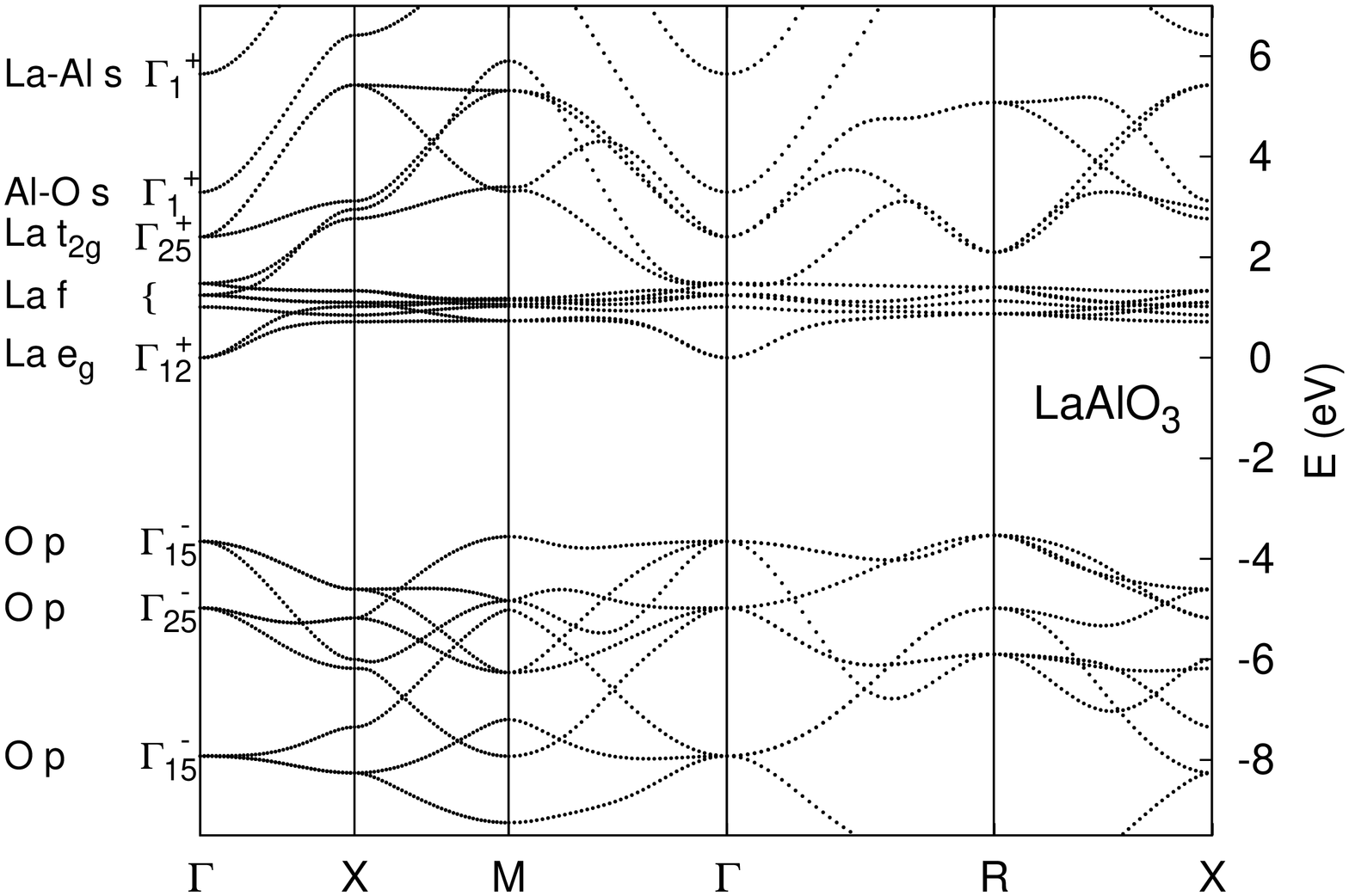}
\caption{Bulk band structure of (top) cubic SrTiO$_3$ and (bottom) cubic LaAlO$_3$ calculated without spin-orbit interaction. The states at $\Gamma$ are labeled by the symmetry representation and by the dominant atomic character. The La f states have symmetry $\Gamma_2^-$, $\Gamma_{25}^-$, $\Gamma_{25}^-$ with increasing energy. Notice that only the $\Gamma_{25}^+$ Ti $t_{2g}$ bands are considered to calculate the energy subbands at the  LaAlO$_3$/SrTiO$_3$ interface (see text).}
\label{fig:symmetries}
\end{center}
\end{figure}

The {\bf k}$\cdot${\bf p} envelope function method requires a description of the band edges of the constituents and their offset. By appropriate matching of the wavefunctions at the interface, one can calculate the energy levels. Both SrTiO$_3$ and LaAlO$_3$ are insulators, with  experimental band gaps of 3.2 and 5.6 eV respectively.\cite{van2001bulk,huijben2009structure} In Fig.~\ref{fig:symmetries}, we show the energy bands of the two constituents, calculated  as described in section~\ref{sec:DFT calculations}, labeled with the symmetry and atomic character at $\Gamma$. Based on the position and symmetry of the bands in the two materials, we argue that a description of the SrTiO$_3$ band edge only is sufficient. 
In fact, the conduction band of SrTiO$_3$ has a minimum at the $\Gamma$ point constituted by the 6-fold degenerate (including spin)  Ti $t_{2g}$ bands. The $t_{2g}$ bands are a subset of the $d$ bands, $d_{xy}, d_{yz}$ 
and $d_{zx}$, split by the crystal field from the other $d$ bands, the $e_g$ states which lie $\sim 2$ eV higher.
In the cubic perovskite structure, the $t_{2g}$ bands at $\Gamma$ transform as $\Gamma_{25}^+$.\cite{dresselhaus2008group} These bands should be matched to states of the same symmetry in the LaAlO$_3$ layer. However, the lowest $\Gamma_{25}^+$ states in the LaAlO$_3$  are located well above the band gap and since the band-offset is expected to be of type I,~\cite{popovic2008origin} even further above the $t_{2g}$ states in SrTiO$_3$. Moreover, in  LaAlO$_3$, the states of this symmetry 
are localized on La, namely on a different crystalline position than the Ti in the SrTiO$_3$, with consequent  small overlap.  Therefore, we  can safely neglect a spreading of the Ti $t_{2g}$ states in the LaAlO$_3$ and assume an infinite potential barrier at the interface. This greatly simplifies the problem, avoiding explicit matching of the wavefunctions in the two materials.

From now on we focus on the 6-fold degenerate $t_{2g}$ states in SrTiO$_3$, which have $\Gamma_{25}^+$ symmetry like the $p$-states in  III-V semiconductors with zincblende structure.\cite{dresselhaus2008group} Note that, due to the crystal field, the orbital momentum $l=2$ of the $d$-states is no longer a good quantum number. 
The spin-orbit splitting will be considered as a perturbation on this level structure and will split the 6-fold degeneracy into a 4-fold $J=3/2$ multiplet of symmetry $\Gamma_{8}^+$ and a 2-fold $J=1/2$ multiplet of symmetry  $\Gamma_{7}^+$, exactly as in the valence band of III-V semiconductors.\cite{dresselhaus2008group}

Following Ref.~\onlinecite{bistritzer2011electronic} we describe the bulk Ti $t_{2g}$ bands around $\Gamma$ by means of a  {\bf k}$\cdot${\bf p} Hamiltonian depending on 3 effective mass parameters $L,M,N$, the spin-orbit splitting $\Delta_{SO}$ and the low temperature tetragonal distortion energy splitting $\Delta_T$:
\begin{equation}\label{eq:Htot}
 H=H_{cubic}(L,M,N)+H_{SO}(\Delta_{SO})+H_T(\Delta_T).
\end{equation}
We choose as basis functions the six $\bm{k}=\bm{0}$ $t_{2g}$ states $\ket{X\uparrow},\ket{Y\uparrow},\ket{Z\uparrow},\ket{X\downarrow},\ket{Y\downarrow},$\mbox{$\ket{Z\downarrow}$} corresponding to $d_{yz}, d_{zx}$ and $d_{xy}$ with both spin up and spin down, although $H_{cubic}$ and $H_T$ do not depend on spin. 
\begin{equation}
  H_{cubic}=\left(
 \begin{array}{cc}
  H_{\Gamma_{25}^+} & \bm{0} \\
  \bm{0} & H_{\Gamma_{25}^+}
 \end{array}
 \right)
 \end{equation}
where  $H_{\Gamma_{25}^+}$ describes the $\Gamma_{25}^+$ bands in a cubic crystal
\begin{widetext}
\begin{equation}\label{eq:cubicH}
 H_{\Gamma_{25}^+}=\left(
 \begin{array}{ccc}
 Lk_x^2+M(k_y^2+k_z^2)&Nk_xk_y & Nk_xk_z\\
  Nk_xk_y & Lk_y^2+M(k_x^2+k_z^2)& Nk_yk_z \\
  Nk_xk_z & Nk_yk_z & Lk_z^2+M(k_x^2+k_y^2)
 \end{array}
 \right).
\end{equation}
\end{widetext}
In this Hamiltonian the interaction with remote bands is taken into account by the effective mass parameters $L, M$ and $N$. $N$ quantifies the coupling between the $t_{2g}$ states via states with other symmetries.

At a temperature $T=110$~K, SrTiO$_3$ undergoes a structural phase transition to a tetragonal symmetry. At $T=4.2~$K the TiO octahedra have rotated $2.1^\circ$ around the tetragonal axis.\cite{mattheiss1972effect} We include this distortion in our model because we want to compare our results to low temperature transport experiments at $T=4.2~$K.  For the heterostructures, we choose   the tetragonal axis along the growth axis which we call the $z$ axis throughout this article. $H_T$ has only two 
non-zero matrix elements $\bra{Z\uparrow}V\ket{Z\uparrow}=\bra{Z\downarrow}V\ket{Z\downarrow}=\Delta_T$ that, at $\Gamma$, shift the $\ket{Z}$ band  $\Delta_T$ above  $\ket{X}$ and $\ket{Y}$. 

The effect of spin-orbit interaction on the $t_{2g}$ bands is described  only by the splitting $\Delta_{SO}$ at $\Gamma$ between the 4-fold $\Gamma_{8}^+$ and 2-fold $\Gamma_{7}^+$ level:
\begin{equation}\label{eq:HsoXYZ}
H_{SO}=\frac{\Delta_{SO}}{3}\left(
\begin{array}{cccccc}
 0 & i & 0 & 0 & 0 & -1 \\
 -i & 0 & 0 & 0 & 0 & i\\
 0 & 0 & 0 & 1 & -i & 0 \\
 0 & 0 & 1 & 0 & -i & 0 \\
 0 & 0 & i & i & 0 & 0 \\
 -1 & -i & 0 & 0 & 0 & 0
\end{array}
\right).
\end{equation}
As the spin-orbit splitting  is important near $\Gamma$, it is convenient to introduce the $\ket{J,m_j}$ basis in which $H_{SO}$ is diagonal
\begin{equation}\label{eq:ubasis}
\begin{aligned}
u_1^8(\bm{r})=&\ket{\frac{3}{2},\frac{3}{2}} =\frac{1}{\sqrt2}\ket{X + i Y\uparrow} \\
u_2^8(\bm{r})=&\ket{\frac{3}{2},-\frac{1}{2}}=\frac{1}{\sqrt{6}}\ket{X -iY\uparrow} + \sqrt{\frac{2}{3}}\ket{Z\downarrow}\\ 
u_3^8(\bm{r})=&\ket{\frac{3}{2},\frac{1}{2}} =\frac{i}{\sqrt6}\ket{X +iY\downarrow} - i \sqrt{\frac{2}{3}}\ket{Z\uparrow}\\ 
u_4^8(\bm{r})=&\ket{\frac{3}{2},-\frac{3}{2}}=\frac{i}{\sqrt2}\ket{X - i Y\downarrow} \\ 
u_5^7(\bm{r})=&\ket{\frac{1}{2},\frac{1}{2}} =\frac{1}{\sqrt3}\ket{X+iY\downarrow}+\frac{1}{\sqrt3}\ket{Z\uparrow} \\ 
u_6^7(\bm{r})=&\ket{\frac{1}{2},-\frac{1}{2}}=-\frac{i}{\sqrt3}\ket{X-iY\uparrow}+\frac{i}{\sqrt3}\ket{Z\downarrow}.
\end{aligned}
\end{equation}
In this basis 
\begin{equation}
H_{SO}=\frac{\Delta_{SO}}{3}\, \text{diag}[-1,-1,-1,-1,2,2]
\end{equation}
which lowers the $J=3/2$ multiplet by $\Delta_{SO}/3$ and raises the $J=1/2$ multiplet by $2\Delta_{SO}/3$. In the following, for simplicity, we take the origin of energy at the $J=3/2$ multiplet which is the minimum of the conduction band. Since the lowest states are almost purely eigenstates of the total angular momentum $J$, it is convenient to write  the matrix H explicitly in the $\ket{J,m_j}$ basis of Eq.\eqref{eq:ubasis}:
\begin{equation}\label{eq:HtotJ}
H=\left(
\begin{array}{cccccc}
p 										& b								 				& -ia					 	 							& 0 					& a/\sqrt2				 							& -i\sqrt 2 b\\
b^\dagger 						& q								 				& 0 						 							& ia 					& \sqrt{\frac{3}{2}} a^\dagger 	& c^\dagger \\
ia^\dagger						& 0 					     				& q						 	 							& b			 	  	& c^\dagger 														& \sqrt{\frac{3}{2}} a \\
 0 										& -ia^\dagger			 				& b^\dagger		 	 							& p				  	& -i\sqrt2 b^\dagger						& a^\dagger/\sqrt 2 \\
a^\dagger/\sqrt2	 		& \sqrt{\frac{3}{2}}	  a & c 													& i \sqrt2 b	& r															& 0 \\
i \sqrt{2} b^\dagger	& c 											& \sqrt{\frac{3}{2}}a^\dagger & a/\sqrt2 		& 0 														& r
\end{array}
\right)
\end{equation}
with
\begin{equation}
\begin{aligned}
a=&\frac{1}{\sqrt{3}}N(k_x-ik_y)k_z\\ 
b=&\frac{1}{2\sqrt{3}}(L-M)(k_x^2-k_y^2)-\frac{1}{\sqrt{3}}iNk_xk_y \\
c=&\frac{1}{3\sqrt2}i(L-M)(k_x^2+k_y^2-2k_z^2)-\frac{\sqrt2}{3}i\Delta_T \\
p=&\frac{1}{2}(L+M)(k_x^2+k_y^2)+Mk_z^2  \\
q=&\frac{1}{6}(L+5M)(k_x^2+k_y^2)+\frac{1}{3}(2L+M)k_z^2+\frac{2}{3}\Delta_T \\
r=&\frac{1}{3}(L+2M)(k_x^2+k_y^2+k_z^2)+\Delta_{SO}+\frac{1}{3}\Delta_T.\\
\end{aligned}
\end{equation}
The effective mass parameters $L,M$ and $N$ and the energy splittings $\Delta_T$ and $\Delta_{SO}$ have to be determined by  experiments or calculations.
As discussed in detail in Ref.~\onlinecite{bistritzer2011electronic} the experimental data for SrTiO$_3$ is rather scarce and contradictory and therefore we have performed DFT calculations which also allow to study each term in the Hamiltonian separately.  

\section{Band structure parameters}\label{sec:DFT calculations}
\subsection{DFT calculations}
We perform first-principles calculations in the framework of 
DFT (Refs.~\onlinecite{hohenberg1964inhomogeneous,PhysRev.140.A1133}) employing the 
Perdew-Burke-Ernzerhof (PBE) generalized gradient approximation 
(GGA).\cite{perdew1996generalized}
The projector augmented wave (PAW) 
method\cite{blochl1994projector,kresse1999ultrasoft} is applied,
as implemented in the Vienna {\it ab initio} simulation program 
(VASP).\cite{PhysRevB.54.11169,kresse1996efficiency}
We use standard PAW data sets as provided with the VASP package,
which have for Sr a frozen [Ar]$3d^{10}$ core, for Ti and Al a frozen [Ne] core, 
and for La a frozen [Kr]$4d^{10}$ core. The La data set includes $f$-channel augmentation.
Calculations with a harder data set (smaller PAW core radii) for oxygen, confirm our results.
VASP uses spinors for calculations with spin-orbit coupling in the Kohn-Sham (KS) Hamiltonian.

A kinetic energy cutoff of 400~eV is employed for the plane wave 
expansion of the KS orbitals.
The calculations are done with a $\Gamma$-centered $12\times12\times12$ 
$\bm{k}$-mesh.\cite{monkhorst1976special}
We use experimental lattice parameters at room temperature: 
$a=3.905$~\AA\ for SrTiO$_3$ (Ref.~\onlinecite{okazaki1973lattice}) and 
$a=3.789$~\AA\ for LaAlO$_3$.\cite{nakatsuka2005cubic}
For body-centered tetragonal SrTiO$_3$ we use $a=3.896$~\AA\ and $c=3.899$~\AA{}.\cite{okazaki1973lattice}

To find the character of the bands, we project the KS states onto
spherical harmonics, within a radius $r_{\rm p}$. For SrTiO$_3$ we choose 
$r_{\rm p} \approx 1.2$~\AA\ for all elements.

The resulting band structures for cubic SrTiO$_3$ and cubic LaAlO$_3$ 
are shown in Fig.~\ref{fig:symmetries}.

\subsection{Fit effective mass parameters}
\begin{figure*}[ht!]
\begin{center}
\includegraphics[angle=0,width=1.0\textwidth]{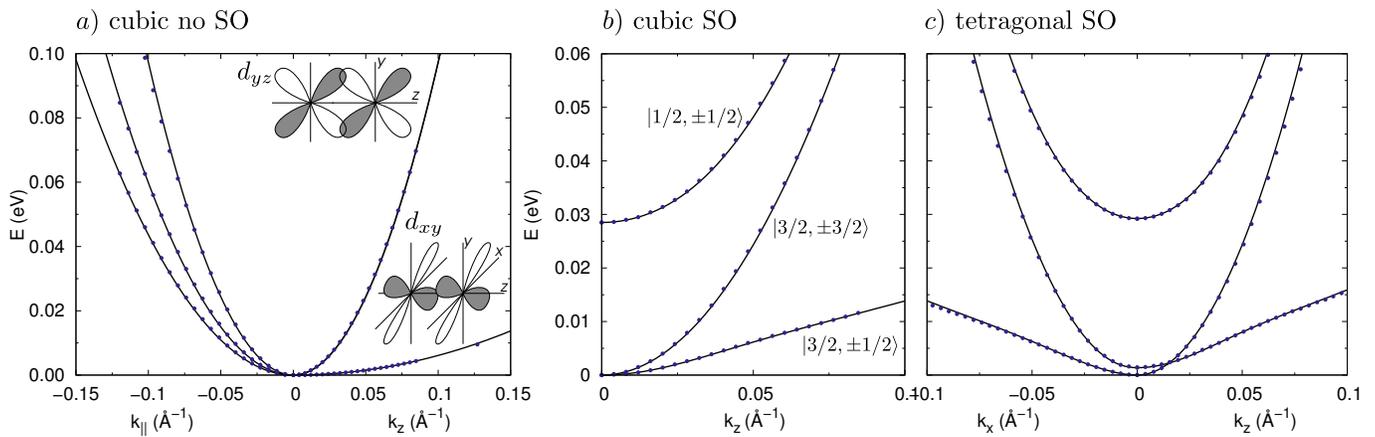}
\caption{Dispersions of the $t_{2g}$ bands in cubic SrTiO$_3$ according to DFT calculations (dots) and to the {\bf k}$\cdot${\bf p} model (lines) with the parameters  given in Table~\ref{table:parameters}. (a) cubic SrTiO$_3$ without SO ($H=H_{cubic}$) along $k_z$ and $k_{||}=(\kappa,\kappa,0)$. Along $k_z$ the steepest band is 4-fold degenerate. The inset shows the $d_{yz}$ orbitals (upper) and $d_{xy}$ orbitals (lower). The three distinct bands along $k_{||}$ indicate $N\neq0$. (b) cubic SrTiO$_3$ with SO ($H=H_{cubic}+H_{SO}$). Note that the bands are shifted by $\Delta_{SO}/3$ so that the $j=3/2$ multiplet edge is at $E=0$. (c) tetragonally distorted SrTiO$_3$ with SO ($H=H_{cubic}+H_{SO}+H_{\Delta_T}$). Note that $k_x$ and $k_z$ are no longer equivalent. 
}
\label{fig:DFTfits}
\end{center}
\end{figure*}

The energy band dispersion given by Eq.~\ref{eq:cubicH} along the three high symmetry lines $k=(\kappa,0,0)$, $k_{||}=(\kappa,\kappa,0)$ and $k=(\kappa,\kappa,\kappa)$ can be written analytically in terms of $L,M$ and $N$. These effective mass parameters are then determined by fitting to the DFT band structure of cubic SrTiO$_3$ without spin-orbit coupling. The $k_x$-, $k_y$- and $k_z$-directions are equivalent because of the cubic symmetry. The resulting values for $L,M$ and $N$ are listed in Table~\ref{table:parameters}.
\begin{table}[]
\centering
\begin{tabular}{c|l|l} 
\hline
\hline
{\bf k}$\cdot${\bf p} parameters & numerical value & calculation\\
\hline
$L$ & $0.6104(2)~\text{eV\AA}^2$  & cubic no SO\\
$M$ & $9.73(2)~\text{eV\AA}^2$ & cubic no SO \\
$N$ & $-1.616(4)~\text{eV\AA}^2$ & cubic no SO\\
$\Delta_{SO}$ & $28.5~$meV & cubic with SO\\
$\Delta_T$ & $2.1~$meV & tetragonal no SO\\
\hline
\hline
\end{tabular}
\caption{{\bf k}$\cdot${\bf p} parameters found by fitting the DFT band structure of SrTiO$_3$. } 
\label{table:parameters}
\end{table}
Fig.~\ref{fig:DFTfits}a shows that the {\bf k}$\cdot${\bf p} model accurately reproduces the DFT calculations at least up to $k=0.15$~\AA{}$^{-1}$. Along $k_z$ there is one very flat band with $m^*=6.24~m_0$. This unusual heavy mass originates from the fact that the $\ket{Z}$ state lies in the $xy$-plane and does not extend along the $z$ direction, as shown in the inset of Fig.~\ref{fig:DFTfits}a. Along $k_{||}$ we find three distinct dispersions given by $1/2(L+M+N)k^2$, $1/2(L+M-N)k^2$ and $Mk^2$ which indicate that  $N\neq0$, in contrast to what has been used in Ref.~\onlinecite{khalsa2012theory}.

\subsection{Calculation of  $\Delta_{SO}$ and $\Delta_T$}

In this Section we discuss how the inclusion of SO coupling and of a tetragonal distortion modify the band structure of Fig.~\ref{fig:DFTfits}a.
To determine $\Delta_{SO}$ we calculate the DFT energy band structure of cubic SrTiO$_3$ including spin-orbit coupling. We find $\Delta_{SO}=28.5~$meV, in agreement with other calculations.\cite{zhong2012theory} 
Fig.~\ref{fig:DFTfits}b shows the resulting 4-fold  $J=3/2$ states and the 2-fold $J=1/2$  split-off bands at $\Gamma$. Although the latter is located $\Delta_{SO}$ above $J=3/2$, it affects the dispersion of the  $\ket{3/2, \pm 1/2}$ band to which it is coupled at finite $k_z$. 
Along $k_z$, the $\ket{3/2, \pm 3/2}$ and the $\ket{3/2, \pm 1/2}$ are the light and heavy mass electrons respectively. This is opposite to the well-studied case of $p$-type valence bands in III-V semiconductors, which can be easily understood by noting that while  the $p_z$ lobe extends mainly in the $z$ direction, the $\ket{Z}$ state extends only in the $xy$-plane. 
Notice that the spin-orbit interaction makes the effective heavy electron mass much lighter, bringing it from $m^*=6.24~m_0$ to $m^*=1.17~m_0$ and with an almost linear dispersion at larger $k_z$. 

The tetragonal distortion energy splitting $\Delta_T$ is determined by calculating the band structure of the tetragonally distorted SrTiO$_3$ without spin-orbit interaction. We find $\Delta_T=2.1~$meV, close to the values found in Refs.~\onlinecite{khalsa2012theory} and \onlinecite{el2013structural}. The tetragonal distortion breaks the symmetry between $k_x$ and $k_z$ as is shown for the case of tetragonally distorted SrTiO$_3$ including SO in Fig.~\ref{fig:DFTfits}c. 

In summary, the {\bf k}$\cdot${\bf p} model is in excellent agreement with the DFT calculations at least up to  $k\sim0.1$~\AA{}$^{-1}$, which is the relevant range for confinement over lengths 
larger than $\sim 40$~\AA.

\section{Quantum well subbands}
\label{sec:QWsub}
In the previous section we have shown that the {\bf k}$\cdot${\bf p} method describes  the conduction band of bulk SrTiO$_3$ very well. A heterostructure leads to a potential profile along the growth direction $V(z)$, leading to quantization of $k_z$. The wavefunction can then be written as
\begin{equation}
 \psi_{k_{||},k_z}(\bm{r})= e^{i\bm{k}_{||}\cdot\bm{r}_{||}} \bm{u(\bm{r})}\cdot \bm{\phi}(z).
\end{equation}
with  $\bm{k}_{||}=(k_x,k_y,0)$, $\bm{r}_{||}=(x,y,0)$ and the basis functions $u_i(\bm{r})$ have the periodicity of the bulk unit cell. The components of $\bm{\phi}(z)$, $\phi_1(z) \ldots \phi_6(z)$ are the envelope functions replacing $e^{ik_z z}$ which  vary slowly on the scale of the unit cell. They can be found by solving the eigenvalue equation: 

\begin{equation}\label{eq:HphiEphi}
 \left\{ H\left( k_z \rightarrow -i\frac{\partial}{\partial z}\right)+V(z) \mathbf{I} \right\} \bm{\phi}(z)=E(k)\bm{\phi}(z)
\end{equation}

From here on we will always use the full Hamiltonian $H$ of Eq.~\eqref{eq:Htot}. We solve this equation by the finite difference method, where we discretize the envelope wavefunction on an equispaced grid in real space.\cite{devries1993first} 
\begin{figure}[h]
\centering
\includegraphics[angle=270,width=8.0cm]{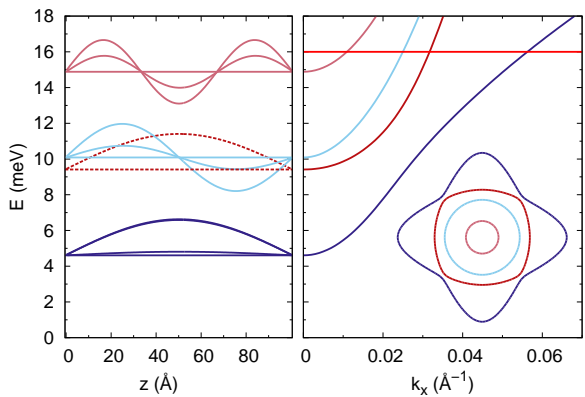}
\caption{(Color online) Left: quantized levels (horizontal lines) and corresponding $\bm{k}=\bm{0}$ envelope functions of a 100 \AA~  SrTiO$_3$ QW. The second subband (dashed lines) has a pure $\ket{3/2, \pm 3/2}$ character.  The envelope  function of the states derived from the  $\ket{3/2, \pm 1/2}$ band (solid lines) are mixed and we plot both  $\ket{3/2, \pm 1/2}$  and $\ket{1/2, \pm 1/2}$ components:  the modulus squared of the $\ket{3/2, \pm 1/2}$ component is 98\%,  88\% and 80\% for increasing energy. Right: in-plane dispersion.  Each subband is plotted with the same color in both panels.  Notice the non-parabolic dispersion due to the strong avoided crossing between the first and second subbands. The horizontal (red) line at $E=16$~meV corresponds to the energy for which the, strongly anisotropic, equal energy contours are shown in the inset.}
\label{fig:QW}
\end{figure}

For illustration, in Fig.~\ref{fig:QW} we show the quantized energy levels in a 100 \AA~ wide SrTiO$_3$ quantum well (QW) with infinite barriers. With increasing energy, the first, third and fourth subbands derive from the heavy $\ket{3/2, \pm 1/2}$ bands and have a mixed $\ket{3/2, \pm 1/2}$ and $\ket{1/2, \pm 1/2}$ character, whereas the second subband has a pure $\ket{3/2, \pm 3/2}$ character.  In the in-plane direction, the heavy  electrons have a light effective mass and vice versa. This leads to strongly non-parabolic dispersion of the in-plane subbands. In particular we see a strong avoided crossing between the first two subbands  at finite in-plane wavevectors.  As a consequence, the equal energy contours are strongly anisotropic and energy dependent, a property which is important for the interpretation of SdH experiments that we will discuss in the next section.

\section{The effect of electric and magnetic fields}
\label{sub:SdHcomp}

The results obtained previously can be used to make a  realistic calculation of the energy levels of different LaAlO$_3$/SrTiO$_3$ heterostructures and   $\delta$ doped SrTiO$_3$. All these systems exhibit low dimensional Shubnikov de Haas oscillations with 1/B periodic oscillations, but with an anomalous amplitude behavior\cite{mccollam2012quantum,kozuka1,kozuka2,caviglia2010,jalan2010,Shalom}. That is, instead of a monotonically increasing amplitude with increasing field, successive oscillations may either be bigger or smaller. Furthermore, several articles mention that the densities obtained from Hall experiments are very different from the ones obtained from the quantum oscillations. These observations indicate multiple subband conduction and in refs. \onlinecite{mccollam2012quantum} and \onlinecite{caviglia2010}, multiple subbands are explicitly mentioned. These results show the need for accurate band structure calculations, such as those we present here. 

Charge neutrality dictates that an interface two-dimensional electron gas requires an equivalent positive charge somewhere in the system.  In the case of a heterostructure, this charge is in the adjacent layers and leads to a constant electric field. This, in turn,  leads  to a potential $V(z)$ that increases linearly with distance from the interface, confining the carriers in the SrTiO$_3$ side of the LaAlO$_3$/SrTiO$_3$ interface. As discussed in section~\ref{sec:model-bulk}, the interface can be modeled by an infinite barrier, leading to a triangular potential
\begin{equation}
\begin{aligned}
 V(z)&=Fz  &\text{if} \quad z\geq0\\
 V(z)&=\infty &\text{if} \quad z<0.
\end{aligned}
\end{equation}
Because of the overall charge neutrality the electric field strength $F$ is directly related to the density of the electron gas at the interface.\cite{khalsa2012theory} As explained above, there is no consensus on the exact origin of the carriers at the LaAlO$_3$/SrTiO$_3$ interface, so that the carrier density  is not fixed. We therefore choose a low carrier density $n\sim 10^{13}$ cm$^{-2}$,  giving rise to $F$ on the order of tenths of meV/\AA{}. These numbers are typical for samples described in refs.~\onlinecite{mccollam2012quantum,caviglia2010,Shalom}.

\begin{figure}[h]
\centering
\includegraphics[angle=270,width=8.0cm]{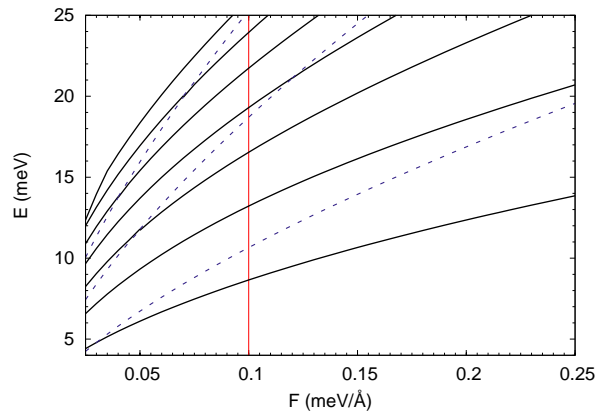}
\caption{(Color online) Dependence of the first 10 quantized levels on $F$. Dashed (blue) lines: $\ket{3/2, \pm 3/2}$ states. The other subbands have a mixed $\ket{3/2, \pm 1/2}$ and $\ket{1/2, \pm 1/2}$ character. Notice the meV spacing in this range of fields. The vertical (red) line is at the field $F=0.1$ meV/\AA~ for which the subband dispersion is presented in Fig.~\ref{fig:3W}.}
\label{fig:EvsF}
\end{figure}

\begin{figure}[h]
\centering
\includegraphics[angle=270,width=8.0cm]{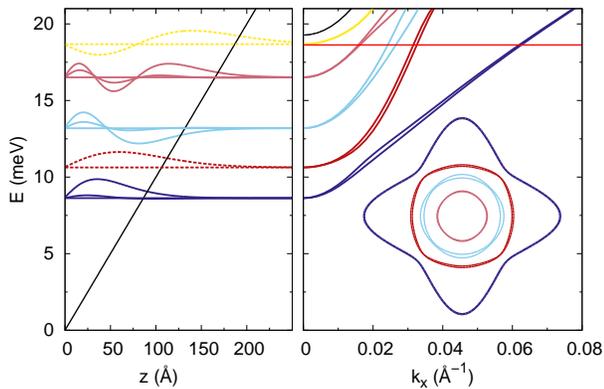}
\caption{(Color online) Left: quantized levels (horizontal lines) and corresponding $\bm{k}=\bm{0}$ envelope functions of a triangular SrTiO$_3$ well with slope $F=0.1$ meV/\AA~ indicated by a solid line.  The second and fifth subbands (dashed lines) have a pure $\ket{3/2, \pm 3/2}$ character. The envelope  function of the states derived from the  $\ket{3/2, \pm 1/2}$ band (solid lines) are mixed and we plot both  $\ket{3/2, \pm 1/2}$  and $\ket{1/2, \pm 1/2}$ components:  the modulus squared of the $\ket{3/2, \pm 1/2}$ component is 98\%,  96\% and 94\% for increasing energy. Right: 
in-plane dispersion. Each subband is plotted with the same color in both panels.  Notice  the almost linear dispersion of the first subband away from $\Gamma$. A small spin splitting is due to the non symmetric confining potential (see text). The (red) horizontal line at $E=18.6~$meV corresponds to the energy for which the equal energy contours are shown in the inset.}
\label{fig:3W}
\end{figure}

In Fig.~\ref{fig:EvsF}, we show the dependence of the first 10 quantized levels on the electric field strength $F$ which represents the slope of the confining triangular well. We show with dashed lines the pure $\ket{3/2 \pm 3/2}$ states. The bands originating from the $\ket{3/2, \pm 1/2}$ bulk bands (solid lines) have a mixed $\ket{3/2, \pm 1/2}$ and $\ket{1/2, \pm 1/2}$ character. The modulus squared of the $\ket{1/2, \pm 1/2}$ is below $10 \%$ for all subbands in the figure.
Notice that, in this range of electric fields, the spacing of the levels is of the order of meV. 
The dispersion is rather complex, with level crossings due to the spin-orbit interaction and the tetragonal distortion.
In Fig.~\ref{fig:3W} we show the envelope functions and subband dispersion calculated for electric field strength $F=0.1~$meV/\AA. We see that the envelope functions at $\bm{k}=\bm{0}$ have the asymmetric shape corresponding to the Airy functions. In this range of fields, the extension of the first subbands is of the order of 100 \AA, which justifies the choice of the envelope function method.
The in-plane dispersion, as for a QW, is strongly non-parabolic and anisotropic, particularly due to the strong avoided crossing between the first and second subband  at finite $\bm{k}$.  Notice that, as already found in the bulk, SO coupling yields a heavy effective mass and an almost linear dispersion  at finite $k$ of the lowest subband.

An additional feature is represented by the small Rashba spin splittings which result from SO coupling in combination with the asymmetric potential.\cite{goldoni1992spin,winkler2003spin,PhysRevLett.78.1335} The spin splitting is strong near avoided band crossings and anisotropic, as can be seen in the inset of Fig.~\ref{fig:3W}. Notice that if the {\bf k}$\cdot${\bf p} coupling $N$ is taken to be zero, the Hamiltonian of Eq.~\eqref{eq:HtotJ} splits into two equivalent $3\times3$ blocks and there is no spin splitting. In other reports $N$ is neglected and a Rashba Hamiltonian is introduced to account for the spin splitting.\cite{zhong2012theory,Khalsaarxiv}

\begin{figure}[h]
\centering
\includegraphics[angle=270,width=8.0cm]{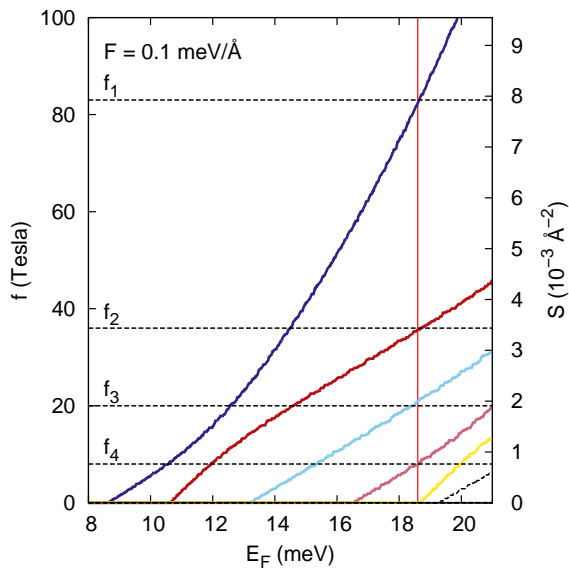}
\caption{(Color online) Dependence  on the Fermi energy of the Fermi surface S (right axis) or of the frequency $f$ of quantum oscillations (left axis)  for all occupied subbands  calculated with $F=0.1$ meV/\AA. The colors correspond to those of Fig.~\ref{fig:3W}. The Fermi surface has been calculated by averaging over the small spin splittings (see Fig.~\ref{fig:3W}). The horizontal dashed lines correspond to the frequencies observed in SdH experiments of Ref.~\onlinecite{mccollam2012quantum} for sample S2: $f_1=83$ Tesla, $f_2=36$ Tesla, $f_3=20$ Tesla, $f_4=8$ Tesla. Notice that at the energy  \mbox{$E_F=18.6$~meV}, indicated by a vertical (red) line, the first four frequencies coincide with the measured values.}
\label{fig:freq} 
\end{figure}

These complex bands and non parabolic in plane dispersions will lead to strongly non linear energy levels in a magnetic field. However, using quasiclassical quantization one can relate the Fermi surface $S$ in $k$-space\cite{kittel1996introduction} directly to the frequency of the quantum oscillations by:  
\begin{equation}
S=\frac{2\pi e f}{\hbar}.
\end{equation}
We can therefore use this relation to compare our calculations to experimental results such as those of Ref.~\onlinecite{mccollam2012quantum}. We calculate the surfaces and corresponding frequencies for various values of $F$ as a function of energy. We average over the Rashba spin-split bands, as these small splittings cannot be resolved in experiments.  In Fig.~\ref{fig:freq} we show the calculated frequencies as a function of the Fermi energy for $F=0.1~$meV/\AA, together with the frequencies that were measured for one of the three samples in Ref.~\onlinecite{mccollam2012quantum} (sample S2). We see that at this value of the electric field, at $E_F=18.6~$meV our calculations are in very good agreement with the experimental values. The electron density for this Fermi energy is $n=7.2 \times 10^{12}$~cm$^{-2}$.
Notice that the large splitting between the first and the second frequency is a general feature that does not depend on the precise values of $F$ and $E_F$. It is a consequence of the almost linear dispersion of the lowest subband at finite $k$.

From the temperature dependence of the SdH oscillations, one can extract an average effective mass at the Fermi energy. As the bands are neither parabolic nor isotropic, the average effective mass at the Fermi energy is difficult to compute for the subbands we have calculated. One way to proceed is to calculate the density of states (DOS) and relate this to the effective mass. For a two-dimensional system with parabolic isotropic bands the DOS is constant
\begin{equation}
\label{eq:effmass}
\text{DOS}_{2D}=\frac{m^*}{2 \pi \hbar^2}
\end{equation}
explicitly counting each spin. We calculate the DOS for each subband  by use of the $k$ space energies on a fine grid, and build a normalized Gaussian  with $0.8~$meV width around each point. The resulting DOS is not constant, as the subbands are neither parabolic nor isotropic. Nevertheless, it represents an average of the subband dispersion over all $k$ points with the same energy, and it can therefore be related  to the energy dependent effective mass by inverting Eq.~\eqref{eq:effmass}:
\begin{equation}
\label{eq:effmassinverted}
m^*(E)=2 \pi \hbar^2 \text{DOS}(E).
\end{equation}

\begin{figure}[h!]
\centering
\includegraphics[angle=270,width=8.0cm]{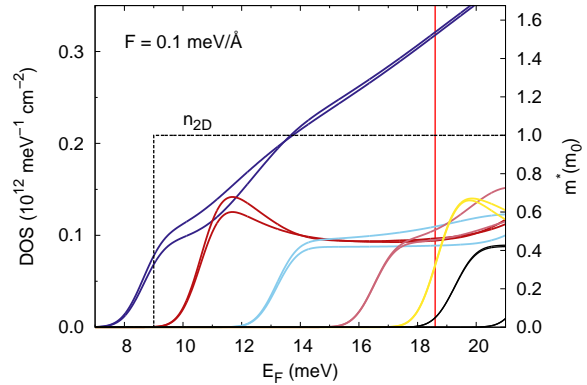}
\caption{(Color online) Dependence on the Fermi energy of the DOS (right axis) and related average effective masses (left axis)  for $F=0.1$~meV/\AA. The colors correspond to those of Fig.~\ref{fig:3W} and  Fig.~\ref{fig:freq}. For reference, the dashed line gives the constant density of states of a free electron ($m^*=m_0$) in two dimensions. Notice that the DOS  deviates  from the free-electron like behavior, particularly for the lowest spin-split subband that grows almost linearly with $E_F$. The vertical (red) line is drawn at $E_F=18.6$~meV.}
\label{fig:masses}
\end{figure}

In Fig.~\ref{fig:masses} we show the DOS and the corresponding effective mass as a function of energy. The precise values of the effective masses are very sensitive to the chosen $E_F$ but the order of heavy and light masses is a robust feature. In Ref.~\onlinecite{mccollam2012quantum} the authors find masses corresponding to the four frequencies reported in Fig.~\ref{fig:freq} of $2.0~m_0$, $0.9~m_0$, $0.9~m_0$, $0.9~m_0$ with decreasing frequency. We also find that the mass of the lowest subband is more than twice that of the following three subbands. For $E_F=18.6$~meV and averaging over the spin-split states we find $1.5~m_0$, $0.5~m_0$, $0.5~m_0$, $0.5~m_0$ which gives a satisfactory agreement. Note that the spin-split bands can have quite different effective masses, for example the effective masses of the third spin-split subband are $m^*=0.42~m_0$ and $m^*=0.52~m_0$.


\section{Summary and perspectives}
\label{sec:S&C}
In summary,  we have calculated the subband structure at the LaAlO$_3$/SrTiO$_3$ interface with the {\bf k}$\cdot${\bf p} envelope function method, using effective mass parameters describing the bulk bandstructure calculated by DFT, leaving no  free parameters for a well defined sample with a known thickness and charge  density. To compare to experimental results, we have assumed a low carrier density resulting in meV spaced subbands and a weak  confining electric field. We have calculated the subband dispersion, density of states,  effective masses and the frequencies of SdH oscillations. We find several occupied, anisotropic, non-parabolic subbands a few meV apart with different, and rather heavy, effective masses as also found experimentally. For an electric field strength $F = 0.1$~meV/\AA, corresponding to  a charge density of $7.2 \times 10^{12}$~cm$^{-2}$, we even find an excellent agreement with specific experimental data.

Our study can be easily extended to consider other structures, since the effective mass {\bf k}$\cdot${\bf p} method allows one to calculate in a very versatile and not too demanding way the effect of structure, layering, strain and composition, as well as  the effect of magnetic and electric fields and doping.

%

\section*{Acknowledgements}
This work is part of the research program of the Stichting voor Fundamenteel Onderzoek der Materie (FOM), which is financially supported by the Nederlandse Organisatie voor Wetenschappelijk Onderzoek (NWO).


\begin{thebibliography}{39}
\expandafter\ifx\csname natexlab\endcsname\relax\def\natexlab#1{#1}\fi
\expandafter\ifx\csname bibnamefont\endcsname\relax
  \def\bibnamefont#1{#1}\fi
\expandafter\ifx\csname bibfnamefont\endcsname\relax
  \def\bibfnamefont#1{#1}\fi
\expandafter\ifx\csname citenamefont\endcsname\relax
  \def\citenamefont#1{#1}\fi
\expandafter\ifx\csname url\endcsname\relax
  \def\url#1{\texttt{#1}}\fi
\expandafter\ifx\csname urlprefix\endcsname\relax\def\urlprefix{URL }\fi

\bibitem[{\citenamefont{Goodenough}(2001)}]{bookgoodenough}
\bibinfo{author}{\bibfnamefont{J.~B.} \bibnamefont{Goodenough}},
  \emph{\bibinfo{title}{Localized to Itinerant Electronic Transition in
  Perovskite Oxides}} (\bibinfo{publisher}{Springer, Berlin}, \bibinfo{year}{2001}).

\bibitem[{\citenamefont{Ohtomo et~al.}(2002)\citenamefont{Ohtomo, Muller,
  Grazul, and Hwang}}]{ohtomo2002artificial}
\bibinfo{author}{\bibfnamefont{A.}~\bibnamefont{Ohtomo}},
  \bibinfo{author}{\bibfnamefont{D.}~\bibnamefont{Muller}},
  \bibinfo{author}{\bibfnamefont{J.}~\bibnamefont{Grazul}}, \bibnamefont{and}
  \bibinfo{author}{\bibfnamefont{H.}~\bibnamefont{Hwang}},
  \bibinfo{journal}{Nature (London) } \textbf{\bibinfo{volume}{419}},
  \bibinfo{pages}{378} (\bibinfo{year}{2002}).

\bibitem[{\citenamefont{Ohtomo and Hwang}(2004)}]{ohtomo2004high}
\bibinfo{author}{\bibfnamefont{A.}~\bibnamefont{Ohtomo}} \bibnamefont{and}
  \bibinfo{author}{\bibfnamefont{H.}~\bibnamefont{Hwang}},
  \bibinfo{journal}{Nature (London) } \textbf{\bibinfo{volume}{427}},
  \bibinfo{pages}{423} (\bibinfo{year}{2004}).

\bibitem[{\citenamefont{Brinkman et~al.}(2007)\citenamefont{Brinkman, Huijben,
  Van~Zalk, Huijben, Zeitler, Maan, Van~der Wiel, Rijnders, Blank, and
  Hilgenkamp}}]{brinkman2007magnetic}
\bibinfo{author}{\bibfnamefont{A.}~\bibnamefont{Brinkman}},
  \bibinfo{author}{\bibfnamefont{M.}~\bibnamefont{Huijben}},
  \bibinfo{author}{\bibfnamefont{M.}~\bibnamefont{Van~Zalk}},
  \bibinfo{author}{\bibfnamefont{J.}~\bibnamefont{Huijben}},
  \bibinfo{author}{\bibfnamefont{U.}~\bibnamefont{Zeitler}},
  \bibinfo{author}{\bibfnamefont{J.}~\bibnamefont{Maan}},
  \bibinfo{author}{\bibfnamefont{W.}~\bibnamefont{Van~der Wiel}},
  \bibinfo{author}{\bibfnamefont{G.}~\bibnamefont{Rijnders}},
  \bibinfo{author}{\bibfnamefont{D.}~\bibnamefont{Blank}}, \bibnamefont{and}
  \bibinfo{author}{\bibfnamefont{H.}~\bibnamefont{Hilgenkamp}},
  \bibinfo{journal}{Nat. Mater.} \textbf{\bibinfo{volume}{6}},
  \bibinfo{pages}{493} (\bibinfo{year}{2007}).

\bibitem[{\citenamefont{Popovi{\'c} et~al.}(2008)\citenamefont{Popovi{\'c},
  Satpathy, and Martin}}]{popovic2008origin}
\bibinfo{author}{\bibfnamefont{Z.S.}~\bibnamefont{Popovi{\'c}}},
  \bibinfo{author}{\bibfnamefont{S.}~\bibnamefont{Satpathy}}, \bibnamefont{and}
  \bibinfo{author}{\bibfnamefont{R.M.}~\bibnamefont{Martin}},
  \bibinfo{journal}{Phys. Rev. Lett.} \textbf{\bibinfo{volume}{101}},
  \bibinfo{pages}{256801} (\bibinfo{year}{2008}).

\bibitem[{\citenamefont{Son et~al.}(2009)\citenamefont{Son, Cho, Lee, Lee, and
  Han}}]{son2009density}
\bibinfo{author}{\bibfnamefont{W.J.}~\bibnamefont{Son}},
  \bibinfo{author}{\bibfnamefont{E.}~\bibnamefont{Cho}},
  \bibinfo{author}{\bibfnamefont{B.}~\bibnamefont{Lee}},
  \bibinfo{author}{\bibfnamefont{J.}~\bibnamefont{Lee}}, \bibnamefont{and}
  \bibinfo{author}{\bibfnamefont{S.}~\bibnamefont{Han}},
  \bibinfo{journal}{Phys. Rev. B} \textbf{\bibinfo{volume}{79}},
  \bibinfo{pages}{245411} (\bibinfo{year}{2009}).

\bibitem[{\citenamefont{Bastard}(1988)}]{bastard1988wave}
\bibinfo{author}{\bibfnamefont{G.}~\bibnamefont{Bastard}},
  \emph{\bibinfo{title}{Wave Mechanics Applied to Semiconductor
  Heterostructures}} (\bibinfo{publisher}{Les \'Editions de Physique, Paris},
  \bibinfo{year}{1988}).

\bibitem[{\citenamefont{Mannhart and Schlom}(2010)}]{mannhart2010oxide}
\bibinfo{author}{\bibfnamefont{J.}~\bibnamefont{Mannhart}} \bibnamefont{and}
  \bibinfo{author}{\bibfnamefont{D.}~\bibnamefont{Schlom}},
  \bibinfo{journal}{Science} \textbf{\bibinfo{volume}{327}},
  \bibinfo{pages}{1607} (\bibinfo{year}{2010}).

\bibitem[{\citenamefont{Khalsa and MacDonald}(2012)}]{khalsa2012theory}
\bibinfo{author}{\bibfnamefont{G.}~\bibnamefont{Khalsa}} \bibnamefont{and}
  \bibinfo{author}{\bibfnamefont{A.~H.} \bibnamefont{MacDonald}},
  \bibinfo{journal}{Phys. Rev. B.} \textbf{\bibinfo{volume}{86}},
\bibinfo{pages}{125121}  (\bibinfo{year}{2012}).

\bibitem{note} See Figs. 7, 3 and 6 of Ref. \onlinecite{khalsa2012theory}.

\bibitem[{\citenamefont{Santander-Syro
  et~al.}(2011)\citenamefont{Santander-Syro, Copie, Kondo, Fortuna, Pailh\`es,
  Weht, Qiu, Bertran, Nicolaou, Taleb-Ibrahimi et~al.}}]{santander2011two}
\bibinfo{author}{\bibfnamefont{A.~F.} \bibnamefont{Santander-Syro}},
  \bibinfo{author}{\bibfnamefont{O.}~\bibnamefont{Copie}},
  \bibinfo{author}{\bibfnamefont{T.}~\bibnamefont{Kondo}},
  \bibinfo{author}{\bibfnamefont{F.}~\bibnamefont{Fortuna}},
  \bibinfo{author}{\bibfnamefont{S.}~\bibnamefont{Pailh\`es}},
  \bibinfo{author}{\bibfnamefont{R.}~\bibnamefont{Weht}},
  \bibinfo{author}{\bibfnamefont{X.~G.} \bibnamefont{Qiu}},
  \bibinfo{author}{\bibfnamefont{F.}~\bibnamefont{Bertran}},
  \bibinfo{author}{\bibfnamefont{A.}~\bibnamefont{Nicolaou}},
  \bibinfo{author}{\bibfnamefont{A.}~\bibnamefont{Taleb-Ibrahimi}}
  \bibnamefont{{\it et~al.}}, \bibinfo{journal}{Nature (London)}
  \textbf{\bibinfo{volume}{469}}, \bibinfo{pages}{189} (\bibinfo{year}{2011}).

\bibitem[{\citenamefont{Delugas et~al.}(2011)\citenamefont{Delugas, Filippetti,
  Fiorentini, Bilc, Fontaine, and Ghosez}}]{delugas2011spontaneous}
\bibinfo{author}{\bibfnamefont{P.}~\bibnamefont{Delugas}},
  \bibinfo{author}{\bibfnamefont{A.}~\bibnamefont{Filippetti}},
  \bibinfo{author}{\bibfnamefont{V.}~\bibnamefont{Fiorentini}},
  \bibinfo{author}{\bibfnamefont{D.~I.} \bibnamefont{Bilc}},
  \bibinfo{author}{\bibfnamefont{D.}~\bibnamefont{Fontaine}}, \bibnamefont{and}
  \bibinfo{author}{\bibfnamefont{P.}~\bibnamefont{Ghosez}},
  \bibinfo{journal}{Phys. Rev. Lett.} \textbf{\bibinfo{volume}{106}},
  \bibinfo{pages}{166807} (\bibinfo{year}{2011}).

\bibitem[{\citenamefont{Breitschaft et~al.}(2010)\citenamefont{Breitschaft,
  Tinkl, Pavlenko, Paetel, Richter, Kirtley, Liao, Hammerl, Eyert, Kopp
  et~al.}}]{breitschaft2010twodimensional}
\bibinfo{author}{\bibfnamefont{M.}~\bibnamefont{Breitschaft}},
  \bibinfo{author}{\bibfnamefont{V.}~\bibnamefont{Tinkl}},
  \bibinfo{author}{\bibfnamefont{N.}~\bibnamefont{Pavlenko}},
  \bibinfo{author}{\bibfnamefont{S.}~\bibnamefont{Paetel}},
  \bibinfo{author}{\bibfnamefont{C.}~\bibnamefont{Richter}},
  \bibinfo{author}{\bibfnamefont{J.~R.} \bibnamefont{Kirtley}},
  \bibinfo{author}{\bibfnamefont{Y.~C.} \bibnamefont{Liao}},
  \bibinfo{author}{\bibfnamefont{G.}~\bibnamefont{Hammerl}},
  \bibinfo{author}{\bibfnamefont{V.}~\bibnamefont{Eyert}},
  \bibinfo{author}{\bibfnamefont{T.}~\bibnamefont{Kopp}} \bibnamefont{{\it et~al.}},
  \bibinfo{journal}{Phys. Rev. B} \textbf{\bibinfo{volume}{81}},
  \bibinfo{pages}{153414} (\bibinfo{year}{2010}).

\bibitem[{\citenamefont{McCollam et~al.}(2012)\citenamefont{McCollam,
  Wenderich, Kruize, Guduru, Molegraaf, Huijben, Koster, Blank, Rijnders,
  Brinkman et~al.}}]{mccollam2012quantum}
\bibinfo{author}{\bibfnamefont{A.}~\bibnamefont{McCollam}},
  \bibinfo{author}{\bibfnamefont{S.}~\bibnamefont{Wenderich}},
  \bibinfo{author}{\bibfnamefont{M.}~\bibnamefont{Kruize}},
  \bibinfo{author}{\bibfnamefont{V.}~\bibnamefont{Guduru}},
  \bibinfo{author}{\bibfnamefont{H.}~\bibnamefont{Molegraaf}},
  \bibinfo{author}{\bibfnamefont{M.}~\bibnamefont{Huijben}},
  \bibinfo{author}{\bibfnamefont{G.}~\bibnamefont{Koster}},
  \bibinfo{author}{\bibfnamefont{D.}~\bibnamefont{Blank}},
  \bibinfo{author}{\bibfnamefont{G.}~\bibnamefont{Rijnders}},
  \bibinfo{author}{\bibfnamefont{A.}~\bibnamefont{Brinkman}}
  \bibnamefont{{\it et~al.}}, \bibinfo{journal}{arXiv:1207.7003}.

\bibitem[{\citenamefont{McCollam et~al.}(2012)\citenamefont{McCollam,
  Wenderich, Kruize, Guduru, Molegraaf, Huijben, Koster, Blank, Rijnders,
  Brinkman et~al.}}]{caviglia2010}
\bibinfo{author}{\bibfnamefont{A.~D.}~\bibnamefont{Caviglia}},
  \bibinfo{author}{\bibfnamefont{S.}~\bibnamefont{Gariglio}},
  \bibinfo{author}{\bibfnamefont{C.}~\bibnamefont{Cancellieri}},
  \bibinfo{author}{\bibfnamefont{B.}~\bibnamefont{Sac\'ep\'e}},
  \bibinfo{author}{\bibfnamefont{A.}~\bibnamefont{F\^ete}},
  \bibinfo{author}{\bibfnamefont{N.}~\bibnamefont{Reyren}},
  \bibinfo{author}{\bibfnamefont{M.}~\bibnamefont{Gabay}},
  \bibinfo{author}{\bibfnamefont{A.~F.}~\bibnamefont{Morpurgo}},
  \bibinfo{author}{\bibfnamefont{J.-M.}~\bibnamefont{Triscone}},
  \bibinfo{journal}{Phys. Rev. Lett.} \textbf{\bibinfo{volume}{105}},
  \bibinfo{pages}{236802} (\bibinfo{year}{2010}).

\bibitem[{\citenamefont{Zhong et~al.}(2013)\citenamefont{Zhong, T\'oth, and
  Held}}]{kozuka1}
\bibinfo{author}{\bibfnamefont{Y.}~\bibnamefont{Kozuka}},
\bibinfo{author}{\bibfnamefont{M.}~\bibnamefont{Kim}},
\bibinfo{author}{\bibfnamefont{C.}~\bibnamefont{Bell}},
\bibinfo{author}{\bibfnamefont{B.~G.}~\bibnamefont{Kim}},
 \bibinfo{author}{\bibfnamefont{Y.}~\bibnamefont{Hikita}} \bibnamefont{and}
  \bibinfo{author}{\bibfnamefont{H.~Y.}~\bibnamefont{Hwang}},
  \bibinfo{journal}{Nature (London)} \textbf{\bibinfo{volume}{462}},
  \bibinfo{pages}{487} (\bibinfo{year}{2009}).

\bibitem[{\citenamefont{Zhong et~al.}(2013)\citenamefont{Zhong, T\'oth, and
  Held}}]{kozuka2}
\bibinfo{author}{\bibfnamefont{M.}~\bibnamefont{Kim}},
\bibinfo{author}{\bibfnamefont{C.}~\bibnamefont{Bell}},
\bibinfo{author}{\bibfnamefont{Y.}~\bibnamefont{Kozuka}},
\bibinfo{author}{\bibfnamefont{M.}~\bibnamefont{Kurita}},
 \bibinfo{author}{\bibfnamefont{Y.}~\bibnamefont{Hikita}} \bibnamefont{and}
  \bibinfo{author}{\bibfnamefont{H.~Y.}~\bibnamefont{Hwang}},
  \bibinfo{journal}{Phys. Rev. Lett.} \textbf{\bibinfo{volume}{107}},
  \bibinfo{pages}{106801} (\bibinfo{year}{2011}).

\bibitem[{\citenamefont{Zhong et~al.}(2013)\citenamefont{Zhong, T\'oth, and
  Held}}]{jalan2010}
\bibinfo{author}{\bibfnamefont{B.}~\bibnamefont{Jalan}},
\bibinfo{author}{\bibfnamefont{S.}~\bibnamefont{Stemmer}},
\bibinfo{author}{\bibfnamefont{S.}~\bibnamefont{Mack}} \bibnamefont{and}
  \bibinfo{author}{\bibfnamefont{S.~J.}~\bibnamefont{Allen}},
  \bibinfo{journal}{Phys. Rev. B} \textbf{\bibinfo{volume}{82}},
  \bibinfo{pages}{081103} (\bibinfo{year}{2010}).

\bibitem[{\citenamefont{Breckenfeld et~al.}(2013)\citenamefont{Breckenfeld,
  Bronn, Karthik, Damodaran, Lee, Mason, and Martin}}]{PhysRevLett.110.196804}
\bibinfo{author}{\bibfnamefont{E.}~\bibnamefont{Breckenfeld}},
  \bibinfo{author}{\bibfnamefont{N.}~\bibnamefont{Bronn}},
  \bibinfo{author}{\bibfnamefont{J.}~\bibnamefont{Karthik}},
  \bibinfo{author}{\bibfnamefont{A.~R.} \bibnamefont{Damodaran}},
  \bibinfo{author}{\bibfnamefont{S.}~\bibnamefont{Lee}},
  \bibinfo{author}{\bibfnamefont{N.}~\bibnamefont{Mason}}, \bibnamefont{and}
  \bibinfo{author}{\bibfnamefont{L.~W.} \bibnamefont{Martin}},
  \bibinfo{journal}{Phys. Rev. Lett.} \textbf{\bibinfo{volume}{110}},
  \bibinfo{pages}{196804} (\bibinfo{year}{2013}).

\bibitem[{\citenamefont{Zhong et~al.}(2013)\citenamefont{Zhong, T\'oth, and
  Held}}]{zhong2012theory}
\bibinfo{author}{\bibfnamefont{Z.}~\bibnamefont{Zhong}},
  \bibinfo{author}{\bibfnamefont{A.}~\bibnamefont{T\'oth}}, \bibnamefont{and}
  \bibinfo{author}{\bibfnamefont{K.}~\bibnamefont{Held}},
  \bibinfo{journal}{Phys. Rev. B} \textbf{\bibinfo{volume}{87}},
  \bibinfo{pages}{161102} (\bibinfo{year}{2013}).

\bibitem[{\citenamefont{Zhong et~al.}(2013)\citenamefont{Zhong, T\'oth, and
  Held}}]{ueno2008}
\bibinfo{author}{\bibfnamefont{K.}~\bibnamefont{Ueno}},
\bibinfo{author}{\bibfnamefont{S.}~\bibnamefont{Nakamura}},
\bibinfo{author}{\bibfnamefont{H.}~\bibnamefont{Shimotani}},
\bibinfo{author}{\bibfnamefont{A.}~\bibnamefont{Ohtomo}},
\bibinfo{author}{\bibfnamefont{N.}~\bibnamefont{Kimura}},
\bibinfo{author}{\bibfnamefont{T.}~\bibnamefont{Nojima}},
\bibinfo{author}{\bibfnamefont{H.}~\bibnamefont{Aoki}},
\bibinfo{author}{\bibfnamefont{Y.}~\bibnamefont{Iwasa}}, \bibnamefont{and}
  \bibinfo{author}{\bibfnamefont{M.}~\bibnamefont{Kawasaki}},
  \bibinfo{journal}{Nat.~Mater.} \textbf{\bibinfo{volume}{7}},
  \bibinfo{pages}{855} (\bibinfo{year}{2008}).

\bibitem[{\citenamefont{van~Benthem et~al.}(2001)\citenamefont{van~Benthem,
  Els\"asser, and French}}]{van2001bulk}
\bibinfo{author}{\bibfnamefont{K.}~\bibnamefont{van~Benthem}},
  \bibinfo{author}{\bibfnamefont{C.}~\bibnamefont{Els\"asser}}, \bibnamefont{and}
  \bibinfo{author}{\bibfnamefont{R.}~\bibnamefont{French}},
  \bibinfo{journal}{J. App. Phys.} \textbf{\bibinfo{volume}{90}},
  \bibinfo{pages}{6156} (\bibinfo{year}{2001}).

\bibitem[{\citenamefont{Huijben et~al.}(2009)\citenamefont{Huijben, Brinkman,
  Koster, Rijnders, Hilgenkamp, and Blank}}]{huijben2009structure}
\bibinfo{author}{\bibfnamefont{M.}~\bibnamefont{Huijben}},
  \bibinfo{author}{\bibfnamefont{A.}~\bibnamefont{Brinkman}},
  \bibinfo{author}{\bibfnamefont{G.}~\bibnamefont{Koster}},
  \bibinfo{author}{\bibfnamefont{G.}~\bibnamefont{Rijnders}},
  \bibinfo{author}{\bibfnamefont{H.}~\bibnamefont{Hilgenkamp}},
  \bibnamefont{and} \bibinfo{author}{\bibfnamefont{D.}~\bibnamefont{Blank}},
  \bibinfo{journal}{Adv. Mater.} \textbf{\bibinfo{volume}{21}},
  \bibinfo{pages}{1665} (\bibinfo{year}{2009}).

\bibitem[{\citenamefont{Dresselhaus et~al.}(2008)\citenamefont{Dresselhaus,
  Dresselhaus, and Jorio}}]{dresselhaus2008group}
\bibinfo{author}{\bibfnamefont{M.}~\bibnamefont{Dresselhaus}},
  \bibinfo{author}{\bibfnamefont{G.}~\bibnamefont{Dresselhaus}},
  \bibnamefont{and} \bibinfo{author}{\bibfnamefont{A.}~\bibnamefont{Jorio}},
  \emph{\bibinfo{title}{Group Theory: Application to the Physics of Condensed
  Matter}} (\bibinfo{publisher}{Springer, Berlin}, \bibinfo{year}{2008}).

\bibitem[{\citenamefont{Bistritzer et~al.}(2011)\citenamefont{Bistritzer,
  Khalsa, and MacDonald}}]{bistritzer2011electronic}
\bibinfo{author}{\bibfnamefont{R.}~\bibnamefont{Bistritzer}},
  \bibinfo{author}{\bibfnamefont{G.}~\bibnamefont{Khalsa}}, \bibnamefont{and}
  \bibinfo{author}{\bibfnamefont{A.~H.} \bibnamefont{MacDonald}},
  \bibinfo{journal}{Phys. Rev. B} \textbf{\bibinfo{volume}{83}},
  \bibinfo{pages}{115114} (\bibinfo{year}{2011}).

\bibitem[{\citenamefont{Mattheiss}(1972)}]{mattheiss1972effect}
\bibinfo{author}{\bibfnamefont{L.}~\bibnamefont{Mattheiss}},
  \bibinfo{journal}{Phys. Rev. B} \textbf{\bibinfo{volume}{6}},
  \bibinfo{pages}{4740} (\bibinfo{year}{1972}).

\bibitem[{\citenamefont{Hohenberg and Kohn}(1964)}]{hohenberg1964inhomogeneous}
\bibinfo{author}{\bibfnamefont{P.}~\bibnamefont{Hohenberg}} \bibnamefont{and}
  \bibinfo{author}{\bibfnamefont{W.}~\bibnamefont{Kohn}},
  \bibinfo{journal}{Phys. Rev.} \textbf{\bibinfo{volume}{136}},
  \bibinfo{pages}{B864} (\bibinfo{year}{1964}).

\bibitem[{\citenamefont{Kohn and Sham}(1965)}]{PhysRev.140.A1133}
\bibinfo{author}{\bibfnamefont{W.}~\bibnamefont{Kohn}} \bibnamefont{and}
  \bibinfo{author}{\bibfnamefont{L.~J.} \bibnamefont{Sham}},
  \bibinfo{journal}{Phys. Rev.} \textbf{\bibinfo{volume}{140}},
  \bibinfo{pages}{A1133} (\bibinfo{year}{1965}).

\bibitem[{\citenamefont{Perdew et~al.}(1996)\citenamefont{Perdew, Burke, and
  Ernzerhof}}]{perdew1996generalized}
\bibinfo{author}{\bibfnamefont{J.P.}~\bibnamefont{Perdew}},
  \bibinfo{author}{\bibfnamefont{K.}~\bibnamefont{Burke}}, \bibnamefont{and}
  \bibinfo{author}{\bibfnamefont{M.}~\bibnamefont{Ernzerhof}},
  \bibinfo{journal}{Phys. Rev. Lett.} \textbf{\bibinfo{volume}{77}},
  \bibinfo{pages}{3865} (\bibinfo{year}{1996}).

\bibitem[{\citenamefont{Bl{\"o}chl}(1994)}]{blochl1994projector}
\bibinfo{author}{\bibfnamefont{P.E.}~\bibnamefont{Bl{\"o}chl}},
  \bibinfo{journal}{Phys. Rev. B} \textbf{\bibinfo{volume}{50}},
  \bibinfo{pages}{17953} (\bibinfo{year}{1994}).

\bibitem[{\citenamefont{Kresse and Joubert}(1999)}]{kresse1999ultrasoft}
\bibinfo{author}{\bibfnamefont{G.}~\bibnamefont{Kresse}} \bibnamefont{and}
  \bibinfo{author}{\bibfnamefont{D.}~\bibnamefont{Joubert}},
  \bibinfo{journal}{Phys. Rev. B} \textbf{\bibinfo{volume}{59}},
  \bibinfo{pages}{1758} (\bibinfo{year}{1999}).

\bibitem[{\citenamefont{Kresse and Furthm\"uller}(1996)}]{PhysRevB.54.11169}
\bibinfo{author}{\bibfnamefont{G.}~\bibnamefont{Kresse}} \bibnamefont{and}
  \bibinfo{author}{\bibfnamefont{J.}~\bibnamefont{Furthm\"uller}},
  \bibinfo{journal}{Phys. Rev. B} \textbf{\bibinfo{volume}{54}},
  \bibinfo{pages}{11169} (\bibinfo{year}{1996}).

\bibitem[{\citenamefont{Kresse and
  Furthm{\"u}ller}(1996)}]{kresse1996efficiency}
\bibinfo{author}{\bibfnamefont{G.}~\bibnamefont{Kresse}} \bibnamefont{and}
  \bibinfo{author}{\bibfnamefont{J.}~\bibnamefont{Furthm{\"u}ller}},
  \bibinfo{journal}{Comput. Mater. Sci.} \textbf{\bibinfo{volume}{6}},
  \bibinfo{pages}{15} (\bibinfo{year}{1996}).

\bibitem[{\citenamefont{Monkhorst and Pack}(1976)}]{monkhorst1976special}
\bibinfo{author}{\bibfnamefont{H.}~\bibnamefont{Monkhorst}} \bibnamefont{and}
  \bibinfo{author}{\bibfnamefont{J.}~\bibnamefont{Pack}},
  \bibinfo{journal}{Phys. Rev. B} \textbf{\bibinfo{volume}{13}},
  \bibinfo{pages}{5188} (\bibinfo{year}{1976}).

\bibitem[{\citenamefont{Okazaki and Kawaminami}(1973)}]{okazaki1973lattice}
\bibinfo{author}{\bibfnamefont{A.}~\bibnamefont{Okazaki}} \bibnamefont{and}
  \bibinfo{author}{\bibfnamefont{M.}~\bibnamefont{Kawaminami}},
  \bibinfo{journal}{Mat. Res. Bull.} \textbf{\bibinfo{volume}{8}},
  \bibinfo{pages}{545} (\bibinfo{year}{1973}).

\bibitem[{\citenamefont{Nakatsuka et~al.}(2005)\citenamefont{Nakatsuka, Ohtaka,
  Arima, Nakayama, and Mizota}}]{nakatsuka2005cubic}
\bibinfo{author}{\bibfnamefont{A.}~\bibnamefont{Nakatsuka}},
  \bibinfo{author}{\bibfnamefont{O.}~\bibnamefont{Ohtaka}},
  \bibinfo{author}{\bibfnamefont{H.}~\bibnamefont{Arima}},
  \bibinfo{author}{\bibfnamefont{N.}~\bibnamefont{Nakayama}}, \bibnamefont{and}
  \bibinfo{author}{\bibfnamefont{T.}~\bibnamefont{Mizota}},
  \bibinfo{journal}{Acta Crystallogr. Sect. E.}
  \textbf{\bibinfo{volume}{61}}, \bibinfo{pages}{i148} (\bibinfo{year}{2005}).

\bibitem[{\citenamefont{El-Mellouhi et~al.}(2013)\citenamefont{El-Mellouhi,
  Brothers, Lucero, Bulik, and Scuseria}}]{el2013structural}
\bibinfo{author}{\bibfnamefont{F.}~\bibnamefont{El-Mellouhi}},
  \bibinfo{author}{\bibfnamefont{E.~N.} \bibnamefont{Brothers}},
  \bibinfo{author}{\bibfnamefont{M.~J.} \bibnamefont{Lucero}},
  \bibinfo{author}{\bibfnamefont{I.~W.} \bibnamefont{Bulik}}, \bibnamefont{and}
  \bibinfo{author}{\bibfnamefont{G.~E.} \bibnamefont{Scuseria}},
  \bibinfo{journal}{Phys. Rev. B} \textbf{\bibinfo{volume}{87}},
  \bibinfo{pages}{035107} (\bibinfo{year}{2013}).

\bibitem[{\citenamefont{DeVries}(1993)}]{devries1993first}
\bibinfo{author}{\bibfnamefont{P.~L.} \bibnamefont{DeVries}},
  \emph{\bibinfo{title}{A First Course in Computational Physics}}
  (\bibinfo{publisher}{John Wiley and Sons, New York}, \bibinfo{year}{1993}).

\bibitem[{\citenamefont{Goldoni and Fasolino}(1992)}]{Shalom}
\bibinfo{author}{\bibfnamefont{M.}~\bibnamefont{Ben Shalom}},
  \bibinfo{author}{\bibfnamefont{A.} \bibnamefont{Ron}},
  \bibinfo{author}{\bibfnamefont{A.} \bibnamefont{Palevski}} \bibnamefont{and}
  \bibinfo{author}{\bibfnamefont{Y.}~\bibnamefont{Dagan}},
  \bibinfo{journal}{Phys. Rev. Lett.} \textbf{\bibinfo{volume}{105}},
  \bibinfo{pages}{206401} (\bibinfo{year}{2010}).

\bibitem[{\citenamefont{Goldoni and Fasolino}(1992)}]{goldoni1992spin}
\bibinfo{author}{\bibfnamefont{G.}~\bibnamefont{Goldoni}} \bibnamefont{and}
  \bibinfo{author}{\bibfnamefont{A.}~\bibnamefont{Fasolino}},
  \bibinfo{journal}{Phys. Rev. Lett.} \textbf{\bibinfo{volume}{69}},
  \bibinfo{pages}{2567} (\bibinfo{year}{1992}).

\bibitem[{\citenamefont{Winkler}(2003)}]{winkler2003spin}
\bibinfo{author}{\bibfnamefont{R.}~\bibnamefont{Winkler}},
  \emph{\bibinfo{title}{Spin-Orbit Coupling Effects in Two-Dimensional Electron
  and Hole Systems}} (\bibinfo{publisher}{Springer, Berlin},
  \bibinfo{year}{2003}).

\bibitem[{\citenamefont{Nitta et~al.}(1997)\citenamefont{Nitta, Akazaki,
  Takayanagi, and Enoki}}]{PhysRevLett.78.1335}
\bibinfo{author}{\bibfnamefont{J.}~\bibnamefont{Nitta}},
  \bibinfo{author}{\bibfnamefont{T.}~\bibnamefont{Akazaki}},
  \bibinfo{author}{\bibfnamefont{H.}~\bibnamefont{Takayanagi}},
  \bibnamefont{and} \bibinfo{author}{\bibfnamefont{T.}~\bibnamefont{Enoki}},
  \bibinfo{journal}{Phys. Rev. Lett.} \textbf{\bibinfo{volume}{78}},
  \bibinfo{pages}{1335} (\bibinfo{year}{1997}).

\bibitem[{\citenamefont{Khalsa et~al.}(2013)\citenamefont{Khalsa, Lee, and
  H.}}]{Khalsaarxiv}
\bibinfo{author}{\bibfnamefont{G.}~\bibnamefont{Khalsa}},
  \bibinfo{author}{\bibfnamefont{B.}~\bibnamefont{Lee}}, \bibnamefont{and}
  \bibinfo{author}{\bibfnamefont{A. H.} \bibnamefont{MacDonald}},
  \bibinfo{journal}{Phys. Rev. B}  \textbf{\bibinfo{volume}{88}},
  \bibinfo{pages}{041302} (\bibinfo{year}{2013}).

\bibitem[{\citenamefont{Kittel}(1996)}]{kittel1996introduction}
\bibinfo{author}{\bibfnamefont{C.}~\bibnamefont{Kittel}},
  \emph{\bibinfo{title}{Introduction to Solid State Physics}}
  (\bibinfo{publisher}{John Wiley and Sons, New York}, \bibinfo{year}{1996}).



\end{thebibliography}

\end{document}